\documentstyle[aps,prb]{revtex}
\include{psfig}

\preprint{}

\begin{document}

\newcommand{\ds}{\displaystyle}
\newcommand{\sm}{\scriptstyle}
\newcommand{\Eq}[1]{Eq.\ (\ref{#1})}
\newcommand{\fig}[1]{Fig.\ \ref{fig:#1}}

\bibliographystyle{prsty}

\title{\bf Phase dependent multiple Andreev reflections in SNS
interferometers}
\author{J.\ Lantz$^a$, V.\ S.\ Shumeiko$^a$, E.\ Bratus$^b$, and G.\
Wendin$^a$}

\address{$^a$ Department of Microelectronics and Nanoscience,
Chalmers University of Technology
and
G\"{o}teborg University, \\S-41296 G\"{o}teborg, Sweden \\
$^b$ Verkin Institute for Low Temperature Physics and Engineering
\\310164 Kharkov, Ukraine}

\date{\today}
\maketitle

\begin{abstract}
A theory of coherent multiple Andreev reflections (MAR) is developed for
superconductor-normal metal-superconductor (SNS)
interferometers. We consider a Y-shape normal electron beam splitter
connecting two superconducting reservoirs, where the two connection
points to the same superconductor can have different phase. The current
is calculated in the quantum transport regime as a
function of applied voltage and phase difference, $I(V,\phi)$.
MAR in interferometers incorporates two features: interference in the
arms of the splitter, and interplay with Andreev resonances. The latter
feature
yields enhancement of the subgap current and current peaks with
phase-dependent positions and magnitudes. The interference effect leads
to
suppression of the subgap current and complete disappearance of the
current
peaks at $\phi=\pi$. The excess current at large voltage
decreases and changes sign with increasing phase difference.

\end{abstract}
%

\narrowtext
\twocolumn

\section{Introduction}
\label{intro}

\begin{figure}[t]
\begin{center}
\psfig{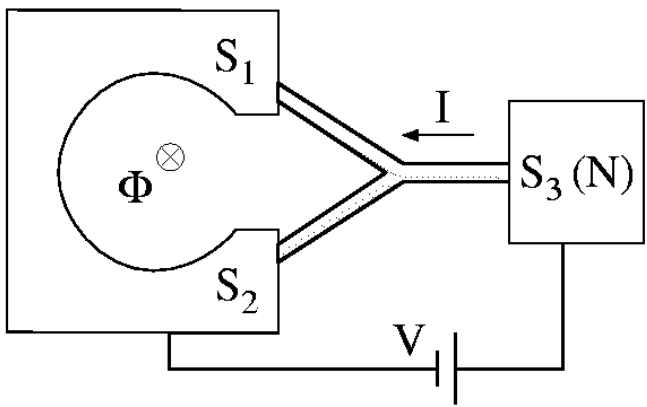}
\end{center}
\caption{Schematic picture of the SNS interferometer, which 
consists of the superconducting loop, to the left, with two electrodes
$S_1$ and $S_2$  that are connected via the Y-branch wave guide to the
right injection electrode $S_3$, which can be superconducting or normal. 
The two left electrodes in the loop and the Y-branch wave guide
constitute an
SNS junction (S$_1$NS$_2$), with bound Andreev states broadened due to
the coupling to the injection electrode S$_3$. 
The phase difference over the S$_1$NS$_2$ junction is controlled by
means of the
magnetic flux through the loop.}\label{fig:setup}
\end{figure}

Mesoscopic circuits with multiterminal electron wave guides have
interesting and useful physical properties.
Figure \ref{fig:setup} shows an example of a circuit where a Y-branch
wave guide acts as a coherent beam splitter for electrons injected from
a normal or superconducting reservoir. Such a circuit with normal
electron
reservoirs (NYN) has been first suggested to test statistical properties
of
the Fermi electrons by measuring current-current correlations in the
arms
of the splitter.\cite{LandauerMartin,Buttiker} This correlation is
negative for
the case of Fermi statistics, which was indeed observed in 
the experiment.\cite{LiuNature} However, the correlations may change the
sign
and become positive (similar to the case of Bose particles) when
the electrons are injected from a superconducting reservoir
(NYS).\cite{Torres}
This effect is related to the Cooper pairing in the superconductor.

Another kind of interference effect, namely a phase dependent
conductance,
has been predicted\cite{Nakano1993,Hekking1993} for a circuit
geometry (SYN) where the arms of the electron beam splitter are
connected to a
superconductor. In this case, the electrons undergo
Andreev reflection from the NS interfaces\cite{Andreev}, picking up
the superconducting phase at the connection points. The phase difference
at
the connection points is created by a supercurrent flowing along the
surface of superconducting electrode, e.g. due to a presence of a 
magnetic field (see Fig. \ref{fig:setup}).
The phase dependence of the conductance of such NS interferometers has
been
observed in a large number of experiments.\cite{Petrashov,Vegvar,review}
%

In this paper, we investigate the properties of 
current-voltage characteristics of SNS interferometers, i.e.
SYS-circuits
where the arms of the Y-branch splitter and the injection lead are
connected to superconducting electrodes, as shown in \fig{setup}. As
is well known, the current between two superconductors at
applied voltage $eV<2\Delta$ ($\Delta$ is the superconducting gap) is
governed by the mechanism of multiple Andreev reflections
(MAR).\cite{KBT} 
The fingerprint of this transport mechanism is 
the subharmonic gap structure (SGS) at
$eV=2\Delta/n$.\cite{Rowel,Bratus1995} 
In SNS interferometers one should expect the SGS to be significantly
modified 
and become sensitive to the superconducting phase difference (see
\fig{setup}). This conclusion is supported
by the following qualitative argument. One may consider the
interferometer as an SNS junction ($S_1NS_2$) which is coupled to the
injection electrode $S_3$
via the Y-branch wave guide. The SNS junction contains bound Andreev
levels which are broadened due to the open connection to the
injection lead. 
The Andreev levels, the energy of which depends on the superconducting
phase difference, produce resonances in the current transport between
the superconducting reservoirs.
These resonances will significantly modify the SGS, similar to the
effect of
superconducting bound states in long SNS junctions.\cite{Ingerman}
However, 
in the SNS interferometers the resonance structures will be phase
dependent.\cite{Wilhelm2000} 

The problem has become particularly interesting due to recent
experiments on a three-terminal SNS interferometer by Kutchinsky et al.,
\cite{Kutchinsky} in which phase dependence of the conductance structures
has been observed. In this experiment, the normal region of the
interferometer was fabricated with diffusive two-dimensional metallic film.
The effect will be more pronounced with a ballistic normal region,
especially in the quantum transport regime,
when the Y-wave guide supports a small amount of conducting
modes. In practice, such devices could be fabricated with etched or gated
ballistic 2D electron gas in a multilayered semiconducting
structure.\cite{Takayanagi1999}
Quantum transport in NS interferometers has been investigated in
Refs.\onlinecite{Zagoskin,Samuelsson,Bagwell}, and the resonance
enhancement of
the two-particle Andreev current has been discussed.
In SNS interferometers the situation
is more complex, since the interplay between the Andreev bound states
and MAR will lead to resonances appearing in all multiparticle
currents.   
The purpose of our study is to develop MAR theory for SNS
interferometers, i.e., to include the
interference effect and the effect of Andreev resonances in the MAR 
calculation scheme\cite{Johansson1999b}
and to analyze the phase dependent current-voltage characteristics
(CVC).

The paper is structured as follows. In section II we present the model
of the interferometer and derive equations for the MAR scattering
amplitudes.
In section III, we present the results of numerical calculation of the
CVC of the interferometer, and in section IV, we present analytical
perturbative analysis of the resonant current structures in the CVC.

\section{Scattering states}

\label{model}

We consider the SNS interferometer shown in \fig{setup}, which consists
of a normal-electron Y-branch quantum wire, fabricated e.g. with etched
or 
gated 2D electron gas, which is connected to two bulk superconducting
electrodes. The superconducting electrode connected to the two arms of
the splitter is ring shaped, so that magnetic flux can be sent through
the ring to induce a superconducting phase difference $\phi$ between the
connection points $S_1$ and $S_2$. 
The current is sent through the interferometer by applying voltage $V$
between the electrode $S_3$ and the ring. 

Our aim is to calculate the 
injected current as a function of the applied voltage and the phase 
difference, $I(V,\phi)$. To this end we shall apply the theory of
coherent multiple Andreev reflections (MAR),
which is based on the calculation of scattering states created by
incoming quasiparticles from the three superconducting 
terminals.\cite{Bratus1995,Averin1995,Bratus1997,Johansson1999b} 
For the sake of simplicity, we adopt a single mode description of the
wires 
and model the splitter by a 3$\times$3 scattering matrix $\hat S$ (see
\Eq{S0} below).

Solving a coherent MAR problem includes the following steps: 
First the recurrence for the scattering amplitudes is formulated by
employing
the boundary conditions at the scatterer and at the NS interfaces. Then
a solution for the recurrence can be constructed, from which the current
is calculated.  In the present case of
a three-terminal junction standard MAR technique cannot be used
directly. 
To solve this problem we will describe the Andreev reflections from
the two interfaces of the interferometer, seen by the injector $S_3$, 
as a complex reflection from a single effective interface; by doing this
we reduce the 3-terminal problem to an equivalent 2-terminal problem, on
which we
can apply standard MAR technique for a 2-terminal junctions developed in
Ref.\onlinecite{Johansson1999b}. 
Reflection from the interferometer (left interface) is then represented
by a non-trivial transfer matrix that contains information about Andreev
states in the interferometer. 
To more clearly understand the effect of the phase difference on the
injection current, we will neglect the length of the normal wires and
assume all NS interfaces to be completely transparent. This will remove
unnecessary complications due to normal-electron resonances
\cite{Johansson1999a} and superconducting
resonances \cite{Ingerman} associated with the Saint-James - de Gennes
bound states.\cite{deGennes1} 
At the same time, the resonant property of the junction as well as the
phase
sensitivity of the current will persist due to the presence of Andreev
states in the
interferometer with energy $E=\pm E_a$,
\begin{equation}
E_a=\Delta\sqrt{1-D\sin^2(\phi/2)}.
\label{Al}
\end{equation}
\begin{figure}[t]
\psfig{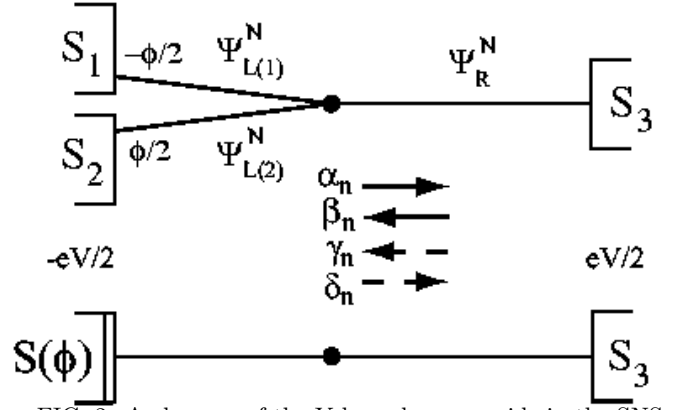}
\caption{A close up of the Y-branch wave guide in the SNS interferometer
(upper figure) and in the effective 2-terminal junction where the left
electrode has phase dependent scattering properties (lower figure). 
We consider a symmetric phase drop over the SNS junction (the electrodes
$S_1$ and $S_2$) and a symmetric voltage drop $eV$ from the loop to the
injection electrode $S_3$. The directions of the scattering amplitudes
$\alpha_n$ and $\beta_n$ (for electrons) and $\gamma_n$ and $\delta_n$
(for holes) are shown with arrows.} \label{fig:setup2}
\end{figure}

In the MAR regime, the time-dependent scattering states consist of a
superposition of
scattered waves with energies $E_n=E+neV$ ($n$ is an integer) shifted
with
respect to the energy of the incoming wave $E$. In each branch of the
wire the wave
functions of the sideband $n$ consists of a
superposition of electron and hole plane waves moving in both
directions,
the corresponding amplitudes being labeled with $\alpha_n,...\delta_n$
according to
\fig{setup2}. In the normal wires, the scattering state wave function
has the form,
\begin{eqnarray}
\Psi_{L(i)}^N & = & \sum_{n=-\infty}^{\infty} \left(\begin{array}{c}
\alpha_{i,n}e^{i k^+_n x}+\beta_{i,n}e^{-i k^+_n x} \\ \gamma_{i,n}e^{i
k^-_n x}+\delta_{i,n}e^{-i k^-_n x}\end{array}\right)e^{-i
(E_n+\frac{eV}2) t} \\
\Psi_{R}^N & = & \sum_{n=-\infty}^{\infty} \left(\begin{array}{c}
\alpha'_{n}e^{i k^+_n x}+\beta'_{n}e^{-i k^+_n x} \\ \gamma'_{n}e^{i
k^-_n x}+\delta'_{n}e^{-i k^-_n x}\end{array}\right)e^{-i
(E_n+\frac{eV}2) t}
\label{ansatzN},
\end{eqnarray}
where $\Psi_{L(i)}^N$ is the wave function in the $i$-th wire at the
left 
side
($i=1,2$), $x$ is a coordinate along the wire, $\Psi_{R}^N$ is the 
wave
function at the right side and
$k^{\pm}_n=\sqrt{2m(E_F\pm E_n)}$ is the normal wave vector ($\hbar=1$).
In the superconducting electrodes, the corresponding wave functions,
including six possible source terms, are:
\begin{eqnarray}
\Psi_{L(1)}^S & = &
e^{-i\sigma_z(\frac{eV}2t-\frac{\phi}4)}\Bigl[(\delta_{1j} \psi_0^+ e^{i
\tilde{k}^+_0 x}+ \delta_{2j} \psi_0^- e^{-i \tilde{k}^-_0 x})e^{-iEt}
\nonumber \\
& + & \sum_{-\infty}^{\infty}(A_n \psi_n^+ e^{-i \tilde{k}^+_n x} + B_n
\psi_n^- e^{i \tilde{k}^-_n x})e^{-iE_nt} \Bigr] \nonumber \\
\Psi_{L(2)}^S & = &
e^{-i\sigma_z(\frac{eV}2t+\frac{\phi}4)}\Bigl[(\delta_{3j} \psi_0^+ e^{i
\tilde{k}^+_0 {x}}+ \delta_{4j} \psi_0^- e^{-i \tilde{k}^-_0
{x}})e^{-iEt} \nonumber \\
& + & \sum_{-\infty}^{\infty}(C_n \psi_n^+ e^{-i \tilde{k}^+_n
{x}} + F_n \psi_n^- e^{i \tilde{k}^-_n {x}})e^{-iE_nt}\Bigr]
\nonumber \\
\Psi_{R}^S & = & e^{i\sigma_z\frac{eV}2t}\Bigl[(\delta_{5j} \psi_0^+
e^{-i \tilde{k}^+_0 x}+ \delta_{6j} \psi_0^- e^{i \tilde{k}^-_0
x})e^{-iEt} \nonumber \\
& + & \sum_{-\infty}^{\infty}(K_n \psi_n^+ e^{i \tilde{k}^+_n x} + L_n
\psi_n^- e^{-i \tilde{k}^-_n x})e^{-iE_nt}\Bigr] \label{ansatzS},
\end{eqnarray}
where $A_n,...L_n$ are constant coefficients and $\psi_n^{\pm}$ are 
elementary solutions to the Bogoliubov-de
Gennes equation \cite{DeGennes1963},
\begin{equation}
\psi_n^{\pm}(E) = {\mbox{min}(1,\sqrt{\Delta/|E_n|})\over
\sqrt{2}}e^{\pm\sigma_z\gamma_n/2}\left(\begin{array}{c}1 \\ \sigma_n
\end{array}\right),
\end{equation}
where
\begin{eqnarray}
e^{\gamma_n} & = &
{|E_n|+\xi_n\over\Delta},\;\;\xi_n=\left\{\begin{array}{ll}
\sqrt{E_n^2-\Delta^2}, & E_n^2>\Delta^2 \\
i\sigma_n\sqrt{\Delta^2-E^2_n}, & E_n^2<\Delta^2 \end{array}\right. ,\\
\sigma_n & = & \mbox{sign}(E_n), \;\;
\tilde{k}_n=\sqrt{2m(E_F\pm\sigma_n\xi_n)}, \nonumber
\end{eqnarray}
and $\sigma_z$ is the Pauli matrix.
The different source terms are distinguished by the index
$j=\{1,2,3,4,5,6\}$
corresponding to injection of an electron/hole from each of the
electrodes.
Injections from the left and from the right generate essentially
different
scattering states and currents, and to keep track of the side of
injection 
we will introduce an additional side index $\nu=\mp 1$
corresponding to injection from the left electrodes and right electrode,
respectively,
\begin{equation}
\begin{array}{lcl} \nu = -1 & \Leftrightarrow & j\in\{1,2,3,4\} \\
\nu = 1 & \Leftrightarrow & j\in\{5,6\}\end{array}.\label{jnu}
\end{equation}

The scattering in the Y-branch splitter is described by a unitary
scattering matrix:
\begin{equation}
\left(\begin{array}{c} \beta_{1,n} \\ \beta_{2,n} \\ \alpha'_{n}
\end{array}\right) = \hat S \left(\begin{array}{c} 
\alpha_{1,n} \\ \alpha_{2,n} \\ \beta'_{n} \end{array}\right),
\label{Scatt}
\end{equation}
where the scattering matrix,
\begin{equation}
\hat S=\left(\begin{array}{ccc} r & d & t \\ d & r & t \\ t & t & r_0
\end{array}\right),
\label{S0}
\end{equation}
is chosen to be symmetric, $S_{31}=S_{32}=t$,
with a real transmission amplitude from right to left, $t>0$, and also
to be independent of energy and thus the same for electrons and holes,
\begin{equation}
\left(\begin{array}{c}
 \delta_{1,n} \\ \delta_{2,n} \\ \gamma'_{n} \end{array}\right) = \hat S
\left(\begin{array}{c} \gamma_{1,n} \\ \gamma_{2,n} \\ \delta'_{n}
\end{array}\right).
\end{equation}
The latter approximation is reasonable since the scale
of the energy dispersion of the scattering matrix is
given by the Fermi energy, while
we are interested in a much smaller energy interval comparable to the
superconducting gap, $E_F\gg\Delta$.
The unitary conditions for $\hat S$ are expressed through the equations,
$1  =  R+D+T$, $1=|r_0|^2+2T$, where $D=|d|^2$, $R=|r|^2$ and $T=t^2$,
and
\begin{equation}
r_0  =  -(r+d)^\ast,\;\;\;
r-d=e^{i\theta'}.\label{unit}
\end{equation}
The transmission coefficient $D$ concerns the transparency of the SNS
junction of the interferometer and controls the position of the Andreev
levels, \Eq{Al}, while the transmission coefficient $T$ describes
coupling of the SNS junction to the injection lead.
 
By means of a canonical transformation the scattering between the
injection electrode and the interferometer can be separated from the
scattering between the arms of the interferometer. To this end we
introduce new scattering amplitudes,
$\alpha_{n}^{\pm}=(\alpha_{1,n}\pm\alpha_{2,n})/\sqrt{2}$,
and similarly for $\beta_{n}, \gamma_{n}$, and 
$\delta_{n}$, and rewrite \Eq{Scatt} on the form,
\begin{eqnarray}
\left(\begin{array}{c}  \beta_n^+ \\ \alpha'_n 
\end{array}\right) = \hat S'\left(\begin{array}{c}  \alpha_n^+ \\
\beta'_n
\end{array}\right), \nonumber \\
\hat S' = \left(\begin{array}{cc} -r_0^\ast & \sqrt{2T} \\ \sqrt{2T} &
r_0 
\end{array}\right),
\label{Scatt2}
\end{eqnarray}
\begin{equation}
\beta_n^- = (r-d)\alpha_n^- 
\label{x1}.
\end{equation}
The scattering equations for holes have a similarly form.
It is clear from \Eq{Scatt2} that the transport through the
interferometer, from the right electrode to the left electrodes, is
independent of the (-)-coefficients. Thus, we can treat the junction as
an effective two-terminal junction with transparency $2T$.

It is convenient for the further calculations to introduce vector
notations,
\begin{eqnarray}
\hat{\alpha}_n^{\pm} &=& \left(\begin{array}{c} \alpha_{n}^{\pm} \\
\beta_{n}^{\pm} \end{array}\right), \;\;
\label{trans}
\end{eqnarray}
and similar for the other scattering amplitudes,
and to rewrite \Eq{Scatt2} through a transfer matrix,
\begin{equation}
\hat{\alpha}_n^+= \hat{T}\hat{\alpha}'_n,
\label{rec3}
\end{equation}
where $\hat{T} = (1/\sqrt{2T})(1-|r_0|e^{-i\sigma_z\rho}\sigma_x)$. 
The same equation also holds for the hole coefficients.
It is convenient to gauge out the reflection phase $\rho$ from
the transfer matrix, which can be done by a transformation of the
scattering amplitudes,
$\hat{\alpha}_n^+,\hat\alpha'_n\rightarrow
e^{i\sigma_z\rho/2}\hat{\alpha}_n^+,
e^{i\sigma_z\rho/2}\hat\alpha'_n $,
whereby
\begin{equation}
\hat{T}\rightarrow (1/\sqrt{2T})(1-|r_0|\sigma_x).
\label{T0}
\end{equation}

We now proceed with a derivation of the recurrence relation and first 
consider the right NS-interface, where we match the wave functions
in \Eq{ansatzN} and (\ref{ansatzS}),
\begin{equation}
\hat{\gamma}'_{n} = \hat V_n^R \hat{\alpha}'_{n-1} +\hat Y_j
\delta_{1\nu}\delta_{0n}\label{rec1}.
\end{equation}
This is the recurrence relation for an ideal SN-interface where the
Andreev reflection amplitude is given by
\begin{equation}
\hat{V}_{n}^R  =  \sigma_n e^{-\sigma_z\gamma_n},  \label{VnR}
\end{equation}
and the source term,
\begin{equation}
\hat Y_{j} = {\sqrt{2}\xi_0\over
\sqrt{|E|}}e^{-\sigma_z\gamma_0/2} \left(\begin{array}c
\delta_{j6} \\ \delta_{j5} \end{array}\right),
\label{pnR}
\end{equation}
corresponds to the injection from the right electrode, $\nu=1$.
Similar relations hold for each of the two SN-interfaces at
the left side. In terms of the vectors $\hat\alpha^\pm$ and
$\hat\gamma^\pm$ introduced in \Eq{trans}, they have the form,
\begin{eqnarray}
\hat{\alpha}_n^+ &=& \sigma_n e^{-\sigma_z\gamma_n}(\hat{\gamma}_{n-1}^+
\cos{\phi\over 2}+\hat{\gamma}_{n-1}^- i\sin{\phi\over 2}) \\
\hat{\alpha}_n^- &=& \sigma_n e^{-\sigma_z\gamma_n}(\hat{\gamma}_{n-1}^+
i\sin{\phi\over 2}+\hat{\gamma}_{n-1}^- \cos{\phi\over 2}),\nonumber
\end{eqnarray}
for $n\neq 0$. For $n=0$ there is also a source term describing
injection 
from the left electrodes ($\nu=-1$). Now the vectors $\hat\alpha^-$ and
$\hat \gamma^-$ can be eliminated by means of \Eq{x1} for electrons and
for holes, which yields
\begin{equation}
\hat{\alpha}_{n}^+ = \hat V_n^L \hat{\gamma}_{n-1}^+ +\hat Y_j
\delta_{-1\nu}\delta_{0n},
\label{rec2}
\end{equation}
where
\begin{equation}
\hat{V}_{n}^L  =  {\sigma_n\over\cos{\phi\over 2}}\left(
e^{-\sigma_z\gamma_n}+\sigma_z(\sigma_xe^{i\sigma_z\theta}-1){\Delta\over
2\xi_n}\sin^2{\phi\over 2}\right),
\label{VnL}
\end{equation}
$\theta=\theta'+\rho$. The source term, $Y_j$, for
$j=1,2$ has the form
\begin{eqnarray}
\hat Y_{j}  &=& {\xi_0\over\sqrt{|E|}}e^{-i\phi/ 4}
\Bigl[e^{-\gamma_0/2} \left(\begin{array}c 1+be^{-\gamma_0} \\
be^{\gamma_0+i\theta} \end{array}\right)\delta_{j1}-\nonumber \\
& & e^{\gamma_0/2} \left(\begin{array}c be^{-\gamma_0-i\theta} \\
1+be^{\gamma_0} \end{array}\right)\delta_{j2}\Bigr],\;\; \nu=-1,
\label{pnL}
\end{eqnarray}
where $b=i\sigma_0(\Delta/2\xi)\tan(\phi/2)$. The source term for the
injection 
cases $j=3,4$ has form similar to \Eq{pnL} with 
$\phi\rightarrow -\phi$.

The matrix $\hat{V}_{n}^L$ describes complete reflection from the
interferometer for energies inside the energy gap, $|E_n|<\Delta$, where
it obeys a standard transfer matrix
equation, $\hat{V}_{n}^L\sigma_z\hat{V}_{n}^{\dagger L}=\sigma_z$.
For $\phi=0$ this is purely Andreev reflection, identical to the one 
described by \Eq{rec1}, which implies that in this case the
interferometer works as a single ideal SN interface. In the general
case, 
$\phi\neq 0$, the reflection consists of both
Andreev and normal reflections, the probability of Andreev reflection
being given by the matrix element $|({V}_n^{L})_{11}|^{-2}\sim
\cos^2(\phi/2)$.
Thus, at $\phi=\pi$ the probability turns to zero, and therefore the 
Andreev transport through the interferometer is blocked. Furthermore,
the transfer matrix $\hat{V}_{n}^L$ contains information about
Andreev bound
states in the interferometer. Assuming for a moment the SNS junction
disconnected from the
injection lead, $T=0$ and considering a stationary version of \Eq{rec2},
$\hat{\alpha}_{n}^+ = \hat V_n^L \hat{\gamma}_{n}^+$, in combination
with
\Eq{T0} we get the solvability condition for these equations on the form
$\mbox{Im}[({V}_n^{L})_{11}-({V}_n^{L})_{12}]=0$,
which gives the Andreev level in \Eq{Al}.

In principle the equations (\ref{rec1}), (\ref{rec2}) and (\ref{rec3})
provide a
complete  set of recurrences for the MAR amplitudes. However, following
Ref.\onlinecite{Johansson1999b}, it is convenient to introduce new 
amplitudes, $\hat c_{n\pm}$, which allow us to get rid of redundant MAR
amplitudes and to unify notation for different injection cases,
\begin{eqnarray}
\hat{c}_{2m+} & = &
\hat{\alpha}_{2m}^+\delta_{-1\nu}+\hat{\gamma}'_{2m}\delta_{1\nu}
\label{c}\\
\hat{c}_{(2m+1)+} & = &
\hat{\gamma}'_{2m+1}\delta_{-1\nu}+\hat{\alpha}_{2m+1}^+\delta_{1\nu}
\nonumber\\ \hat{c}_{2m-} & = &
\hat{\gamma}_{2m-1}^+\delta_{-1\nu}+\hat{\alpha}'_{2m-1}\delta_{1\nu}
\nonumber\\ \hat{c}_{(2m+1)-}& =&\hat{\alpha}'_{2m}
\delta_{-1\nu}+\hat{\gamma}_{2m}^+\delta_{1\nu}.\nonumber
\end{eqnarray}
%
Then the recurrence relations \Eq{rec1}, (\ref{rec2}) and
(\ref{rec3}) can be written on a compact form,
\begin{eqnarray}
\hat{c}_{n+}&=&\hat{U}_n \hat{c}_{n-}+\hat Y_j\delta_{0n}\label{rec}, \\
\hat{c}_{(n+1)-}&=&\hat{T_n}\hat{c}_{n+}\label{recx},
\end{eqnarray}
where
\begin{eqnarray}
\hat{U}_{n} &=&
\delta_{1\mu}\hat{V}_n^{R}+\delta_{-1\mu}\hat{V}_n^{L},\\
\hat{T}_{n} &=& {1\over \sqrt{2T}}(1-\mu|r_0|\sigma_x),
\label{UT}
\end{eqnarray}
and the index $\mu=(-1)^n\nu$ specifies the particular form of the
transfer matrices $\hat{U}_{n}$ and $\hat{T}_{n}$ for different
side bands and for different injection directions.

Equations (\ref{c})-(\ref{UT}) realize a mapping of the MAR problem on
the problem of wave propagation along the energy axis. The wave 
amplitudes for the propagation in the upward and downward directions are 
given by the upper, $c^\uparrow_{n\pm}$, and lower, 
$c^\downarrow_{n\pm}$, components of the vector $\hat{c}_{n\pm}$,
\begin{equation}
\hat{c}_{n\pm} =\left(\begin{array}c c^\uparrow_{n\pm} \\
c^\downarrow_{n\pm}
\end{array}\right).
\label{chat}
\end{equation}
The probability  current flowing 
along the energy axis is defined in the usual way as
\begin{equation}
j_{n\pm}^{p}=|{c}_{n\pm}^{\uparrow}|^2- |{c}_{n\pm}^\downarrow|^2.
\label{probj}
\end{equation}
This probability current is conserved within the superconducting gap, 
$j^p_{n\pm}=$const for $|E_n|<\Delta$, due to
the unitarity properties of the matrices $\hat T_n$ and  $\hat
U_n$. Violation of the unitarity for $\hat U_n$ outside the gap
indicates
leakage of the probability current into the reservoirs. We notice that 
in real space, the
probability current is carried alternatively by electrons and holes.

The charge current through the interferometer is determined by the 
probability currents of all the side bands,
\begin{eqnarray}
I&=&{e\over 2\pi}\sum_\nu 
\int_{|E|>\Delta}dE\ J_\nu(E) n_F(E),\label{Idef} \\
J_\nu(E)&=& {|E|\over\xi} \sum_{j,n=\mbox{\scriptsize
even}}\left( j_{n-}^{p} + j_{n+}^{p}\right)
\label{jdef}.
\end{eqnarray}
In this equation, the spectral current $J_\nu(E)$ consists of the sum
over
the scattering states and all the sidebands at the injection side of the
junction, $\nu$; the factor $|E|/\xi$ is the superconducting density of
states,
and $n_F(E)$ is the Fermi
distribution function within the superconducting reservoirs.

\begin{figure}[t]
\psfig{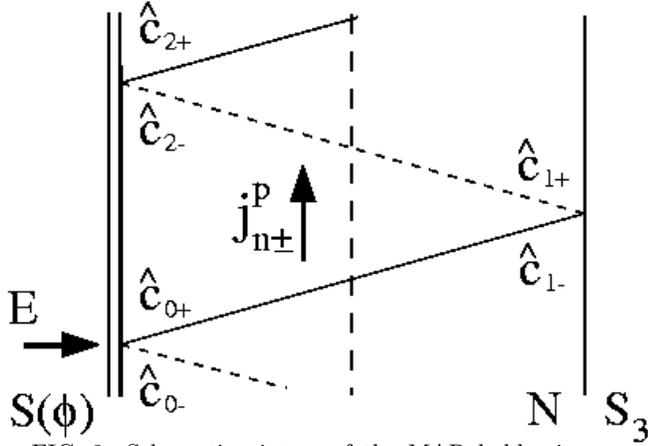}
\caption{Schematic picture of
the MAR ladder in energy space, for an injected quasiparticle at energy
$E$. The interfaces of the junction are shown as vertical lines, where
at the right side there is an ordinary NS interface, while at the left
side
(double line) there is the effective phase dependent SN-interface. The
effective scatterer is shown as a dashed line in the center of the
junction.
The two-component scattering amplitudes $\hat{c}_{n\pm}$ represent up
and down going electrons (solid line) and holes (dashed line). The
corresponding
probability current at level $n$ is $j_{n\pm}^p$.} \label{fig:marfig}
\end{figure}
%

For $n>0$, the free solution of Eqs. (\ref{rec}) and
(\ref{recx}) can be written on the form,
\begin{eqnarray}
\hat{c}_{n-} & = & \hat{M}_{n0}\hat{c}_{0+}, \;\;\;
\hat{c}_{n+}= \hat U_n\hat{c}_{n-} \nonumber \\
\hat{M}_{n0} & = &
\hat{T}_{n-1}\hat{U}_{n-1}\hat{T}_{n-
2}...\hat{U}_{1}\hat{T}_{0},\label{Mnm}
\end{eqnarray}
while the solution for negative $n$, $n<0$, reads
$$\hat{c}_{n+}  =
\hat{M}_{n0}^{-1}\hat{c}_{0-}\;\;, \hat{c}_{n-}= \hat
U_n^{-1}\hat{c}_{n+}.$$
At the injection point $n=0$, the MAR coefficients are found from 
\Eq{rec}, 
\begin{equation}
\hat{c}_{0+}=\hat{U}_{0}\hat{c}_{0-} + \hat Y_j.
\label{rec0}
\end{equation}
This equation imposes two relations between the four unknown 
coefficients, $c_{0\pm}^{\uparrow},\; c_{0\pm}^{\downarrow}$; another
two 
relations are imposed by the boundary conditions at infinity in energy
space
\begin{equation}
\lim_{n\rightarrow\pm\infty}\hat{c}_{n\pm}=0,
\label{lim}
\end{equation}
which fix the ratios $c_{0\pm}^{\uparrow}/ c_{0\pm}^{\downarrow}$.
To implement the boundary conditions, we introduce 
effective reflection coefficients, $r_{n\pm}$, which characterize the 
intensity of the back scattering at specific points of the MAR ladder.
The definition reads,
\begin{eqnarray}
\hat{c}_{n+} &=& c_{n+} ^{\uparrow}\hat r_{n+},\;n\geq 0 \nonumber \\
\hat{c}_{n-}  &=& c_{n} ^{\downarrow}i\sigma_y\hat r_{n-},\;n\leq 0,
\label{rdef}
\end{eqnarray}
where
\begin{equation}
\hat r_{n+}=\left(\begin{array}{c} 1 \\ r_{n+}
\end{array}\right)\;\;
\hat r_{n-}=\left(\begin{array}{c} 1 \\ -r_{n-}
\end{array}\right).
\label{rm}
\end{equation}
With help of the vectors $\hat r_{n\pm}$ the solution in \Eq{Mnm} can be
presented in a compact form (using the equality
$\mbox{det}(\hat{M}_{nm})=1$),
\begin{equation}
\hat{c}_{n+}={(\hat r_{0-}^*,\hat{U}_0^{-1}\hat Y_j)\over
z_{n0}}\;\hat r_{n+} ,\;\;n>0,
\label{c>0}
\end{equation}
where the brackets denote a scalar product of two-component vectors, and
\begin{equation}
z_{n0} = (\hat r_{0-}^*,\hat{U}_0^{-1} \hat{M}_{n0}^{-1}
\hat{U}_n^{-1}\hat r_{n+} ).
\label{z}
\end{equation}
The corresponding expression for negative $n$ is
\begin{eqnarray}
\hat{c}_{n-}&=&-{(\hat r_{0+}^*,\sigma_y\hat Y_j)\over \tilde
z_{0n}}\;\sigma_y\hat r_{n-},\;\;n<0, \label{c<0} \\
\tilde z_{0n}&=& (\hat
r_{0+}^*,\sigma_y
\hat{U}_0 \hat{M}_{0n} \hat{U}_n \sigma_y\hat r_{n-}). \nonumber
\end{eqnarray}
The form of solution in Eqs. (\ref{c>0})- (\ref{c<0}) is
particularly useful for the calculation of the SGS: The essential
processes which contribute to the SGS involve 
transmission through the energy gap region. This transmission
contains resonances which are included in the transfer matrix
$\hat M_{n0}$. The latter is straightforward to calculate because
it does not require excursions to infinity. On the other hand,
the calculation of the quantities $r_{n\pm}$ includes such excursions
which, 
however, go outside the energy gap, and the corresponding recurrences
rapidly converge. 

To calculate the SGS, it is convenient to separate out the current
associated with the scattering across the gap, which is most important
for the
SGS, from the current of thermal excitations which
involves transitions between states below or above the gap.
To this end, we first rewrite
the spectral current in \Eq{jdef} through leakage currents into the
electrodes  defined as, $ j_{n} = j_{n+}^{p}-j_{n-}^{p}$. By using the
equality
$ j_{(n+1)-}^{p} = j_{n+}^{p}$ following from \Eq{recx}, we get
\begin{eqnarray}
J_\nu(E)&=& {|E|\over\xi} \sum_{n,j\in\nu} n \, j_{n,j},
\label{j}
\end{eqnarray}
(here we explicitly write the scattering state index $j$, introduced in
\Eq{ansatzS}).
We note that the conservation of the probability current inside the
energy 
gap implies
that the leakage current is zero inside the gap,
$j_{n,j}=0,\;|E_n|<\Delta$.
Having made this observation, we write the total current in \Eq{jdef} 
on the form,
$$
I={e\over 2\pi}\sum_{j,n=1}^\infty 
\int_{-\infty}^{-\Delta}{dE|E|\over\xi}\; n
\left[ -j_{-n,j} +  \theta(-\Delta-E_n)\, j_{n,j}\right.
$$
\begin{equation}
\left. + \theta(E_n-\Delta)\, j_{n,j} \right] \tanh{E\over 2k_BT} .
\label{Isep}
\end{equation}
The first two terms in this equation correspond to the current of
thermal
excitations: they cancel each other when the temperature approaches zero
due to detailed balance,\cite{Johansson1999b}
\begin{equation}
J_{n,\nu}(E)=J_{-n,\mu}(E_n),
\label{proof}
\end{equation}
for the partial currents defined as
\begin{equation}
J_{n,\nu}(E)= n \;{|E|\over \xi} \sum_{j\in\nu} j_{n,j}.
\label{jnu1}
\end{equation}
Keeping the last term in \Eq{Isep}, 
we get for the current at low temperature, $T\ll\Delta$, the following
equation, 
\begin{eqnarray}
I&=&\sum_{n>0}\theta(eVn-2\Delta)(I_{n}^R+ I_{n}^L),\label{Itot} \\
I_{n}^{\nu}&=&{e\over 2\pi}\int_{\Delta-neV}^{-\Delta}dE\
J_{n,\nu}(E)\tanh\left({E\over 2k_BT}\right) \label{Itotn}.
\end{eqnarray}
The current $J_{n,\nu}(E)$ in \Eq{jnu1} can be 
presented after some algebra on the form which is useful for analytical
study, 
\begin{equation}
J_{n,\nu}(E )=n\;{i_{0,\nu}(E)i_{n,\mu}(E)\over |z_{n0}|^2},
\label{jX}
\end{equation}
where 
%
\begin{eqnarray}
i_{0,\nu}(E)&=&
(\hat r_{0-},[\hat{U}_0^{*-1}\sigma_z\hat{\bar U^{-1}}_0 -
\sigma_z]\hat r_{0-}),\\
i_{n,\nu}(E)&=&
(\hat
r_{n+},[\hat{U}_n^{\dagger-1}\sigma_z\hat{U}_n^{-1}-\sigma_z]r_{n+})
\label{in}\nonumber.
\end{eqnarray}
%
\section{Numerical IVC}
\label{results}
\begin{figure}[b]
\psfig{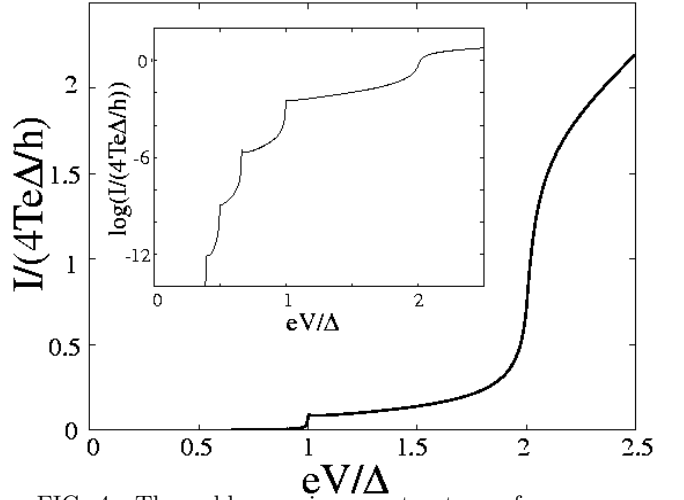} 
\caption{The subharmonic gap structure of a superconducting point
contact with current steps a $eV=2\Delta/n$, $n=1,2,3,...$ (most
visible on the logarithmic diagram). The transmission probability of the 
point contact is $T_{PC}=0.08$, which corresponds to an SNS
interferometer with $\phi=0$ and $2T=0.08$.} \label{fig:iv0}
\end{figure}

In this section, we present the results of numerical calculation of the
current given by \Eq{Itot}, where the spectral current is given by
\Eq{jX}.
In order to numerically evaluate the effective reflection amplitudes
$r_{n\pm}$ we
use the equations, the definitions \Eq{rdef}, using 
\Eq{Mnm} and \Eq{lim}, 
\begin{equation} 
r_{n+}=\lim_{m\rightarrow \infty}{M_{mn}^{\dagger(22)}\over
M_{mn}^{\dagger(21)}},\;\;\;r_{0-}=\lim_{m\rightarrow
-\infty}{M_{0m}^{(12)}\over M_{0m}^{(11)}},
\end{equation} 
which follow from the definitions in \Eq{rdef}, and also from  
\Eq{Mnm} and \Eq{lim}.
For zero phase difference, the current-voltage characteristic (CVC) 
in \fig{iv0}  shows the well known subharmonic gap 
structure for short junctions with current structures at $eV=2\Delta/n$,
corresponding to thresholds for the $n$-particle 
currents\cite{Schrieffer1963,Bratus1995}.
These current structures are associated with the usual gap edge
singularities,\cite{Johansson1999b} and they are most pronounced if the 
transparency $2T$ of the junction is small, $T\ll 1$. 

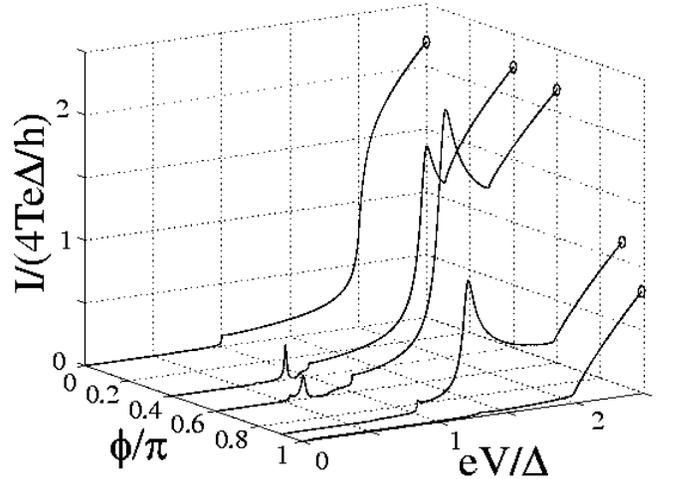
\begin{figure}[b]
\psfig{figure=fig5.epsi,width=8.5cm} 
\caption{The current voltage
characteristics at zero temperature of an interferometer with $2T=0.08$,
$R=0.1$ and $\phi=0$, $\phi=2\pi/5$, $\phi=3\pi/5$, 
$\phi=0.9\pi$ and $\phi=0.99\pi$.} \label{fig:iv1}
\end{figure}

\begin{figure}[b]
\psfig{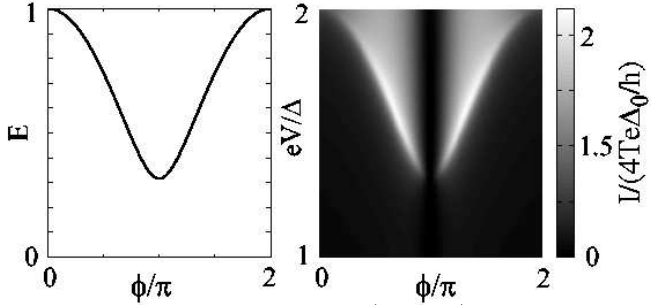} 
\caption{Left: the energy of the (positive) Andreev level for
$D=0.9$, as a function of the phase difference. Right: the current peak
(light
region) at $eV\approx\Delta+E_a(\phi)$ follows the position of the
Andreev 
level. } \label{fig:closeup}
\end{figure}

The CVCs
of the interferometer for different values of the phase difference at 
zero temperature are shown in \fig{iv1}. 
%
%
Increasing the phase difference, the structure changes and the
positions of the current structures can no longer be associated with the
subharmonics of the energy gap. Instead, we observe enhancement of the
subgap 
current and the appearance of large current peaks which move downwards
in 
voltage with increasing phase difference. If the transparency is small,
as in \fig{iv1}, the peak positions reflect the Andreev spectrum of 
\Eq{Al}: the peaks appear at $eV\approx E_a$ and $eV\approx \Delta+E_a$, 
as shown in \fig{closeup}. With further 
increase of the phase difference, the current structures decrease and
completely 
disappear at $\phi=\pi$. At the same time, the excess current at large
voltage becomes large and negative, turning into a deficit current.
The CVC is $2\pi$-periodic and symmetric around $\pi$.
%
%
%
%

\begin{figure}[b]
\psfig{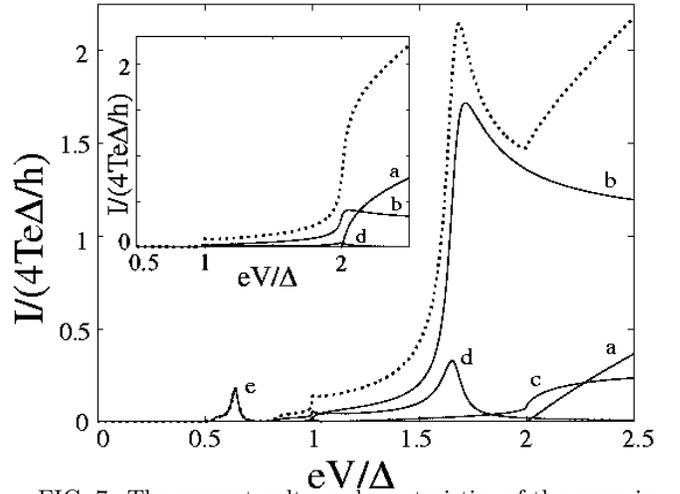} 
\caption{The current voltage
characteristics of the same interferometer as in \fig{iv1} with
$\phi=3\pi/5$.
The total current is shown with dotted line. The single-particle current
(a) is
suppressed compared to the case when $\phi=0$.
The two-particle current (b) of 
quasiparticles injected from the right electrode has a significant onset
at
$eV=\Delta+E_a$, while the two-particle current (c) injected from the
left is
small. The three-particle current (d) has a peak at
$eV\approx \Delta+E_a$; the four-particle current (e) has a
peak at $eV\approx E_a$. The inset shows the
corresponding n-particle CVCs for $\phi=0$.} \label{fig:iv2}
\end{figure}

It is illuminating to plot the individual n-particle currents, given by
\Eq{Itotn}. In \fig{iv2}, the first four n-particle currents are shown 
for $\phi=3\pi/5$ and for the
same interferometer parameters as in \fig{iv1}. Higher order
currents are too small to be visible in the same picture, and their
contribution to the current may be neglected. For comparison, the inset
shows the corresponding CVC when $\phi=0$.
It follows from the plots for $\phi\neq 0$, that the peak at low voltage
originates exclusively 
from the four-particle current. It exists within the voltage interval 
$\Delta/2<eV<\Delta$ and within a certain window of phase
difference. The next peak, at larger voltage, results from a 
combination of two structures: the peak of the three-particle current,
and the overshoot of the onset of the pair current. The onset of a
single-particle 
current at $eV>2\Delta$ is reduced in comparison with the $\phi= 0$
case, which leads to 
significant reduction of the current at large voltage. The reason is
spectral intensity transfer from the continuum to bound Andreev states.
The large deficit
current at high voltage is apparently the result of this reduction
as well as of the above mentioned reduction of the multiparticle
currents related to the 
suppression of Andreev reflections by the 
interferometer. The further analysis shows that 
the single- and three-particle currents (and in general odd $n$-particle
currents) injected from the electrode
$S_3$ and from the electrodes $S_{1,2}$ are identical.
In contrast, the two- and 
four-particle currents possess large structures only for injection from
the electrode $S_3$ while currents injected from the interferometer
electrodes 
$S_{1,2}$ are negligible small. In the next section, we will proceed 
with the interpretation of these features.
%
%

\section{Resonance approximation}
\label{approx}

In the case of weak coupling of the $S_1NS_2$ junction in the
interferometer to the
injection lead $S_3$, $T\ll 1$, it is possible to
perform a perturbative analysis of the n-particle currents in order to
explain 
the origin of the phase dependent current structures in the CVC.
In this limit, 
the current is dominated by the contribution of the Andreev resonances,
which we will consider in the lowest order with respect to the small
parameter $T\ll 1$. We will proceed with a detailed analysis of the four
lowest-order 
n-particle currents in order to explain the major structures in the IVC.

Let us start the discussion with the approximation for the effective
reflection 
amplitudes at specific levels in the MAR ladder, $\hat r_{n\pm}$, in
\Eq{rm}. In the lowest order 
approximation with respect to $T$, these reflection amplitudes are
contributed by single back scattering,
\begin{equation}
\hat r_{n\pm} = \left(\begin{array}{c} 1 \\ \mp \nu(-1)^n r_0
\end{array}\right)+O(T^2).
\end{equation}
Since the denominator $|z_{n0}|^2$ in expression \Eq{jX} for the
spectral
current contains a factor $1/(2T)^n$ which comes from the product of
transfer 
matrices $\hat T_n$ in \Eq{Mnm}, we can neglect all terms except the
zero order one in \Eq{in}, whereby
\begin{eqnarray}
i_{n,\mu}=i_n^L\delta_{-1,\mu}+i_n^R\delta_{1,\mu},\;\;n>0,  \nonumber
\\
i_{n}^{L} = {4|E_n| \over
\Delta^2}{(E_n^2-E_a^2)\over\xi_n\cos^2{\phi\over 2}},\;\;
i_{n}^{R} = {4|E_n| \xi_n \over \Delta^2}.
\label{Iappn}
\end{eqnarray}

The denominator $|z_{n0}|^2$ in \Eq{jX} has to be considered separately
for different currents, $n$, and injection cases, $\nu$. 

\begin{figure}[b]
\psfig{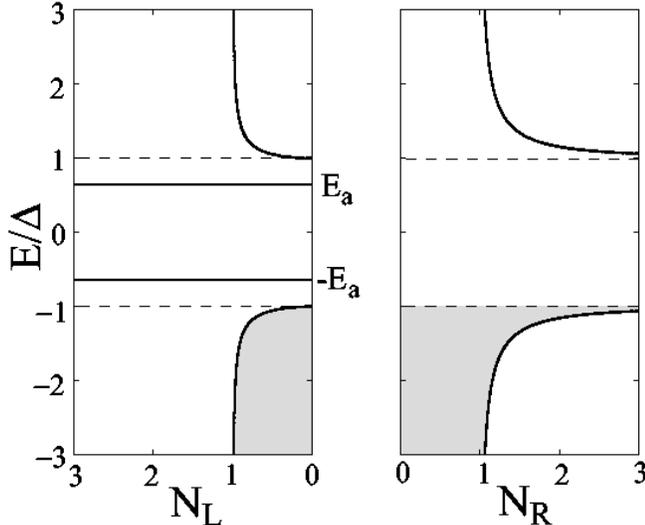} 
\caption{The effective superconducting density of states in the left and
the right electrodes of the SNS interferometer, with $D=0.9$ and
$\phi=3\pi/5$. 
The shaded regions represent the states filled at zero temperature. 
When the phase difference is changed from
zero, the bound Andreev states split from the continuum, and the peaks 
in the density of states at the gap edges are smeared in the left
electrode. 
} \label{fig:dos}
\end{figure}

\begin{figure}[b]
\psfig{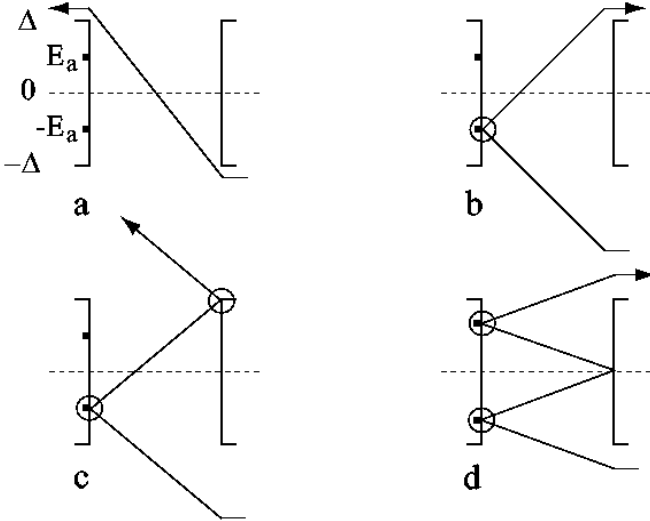} 
\caption{Schematic
pictures of important scattering processes. Resonances are shown as
circles. 
a: Non-resonant single particle current; 
b: resonant pair current from the right side involves one resonant
Andreev reflection, two equivalent resonant paths are possible;
c: resonant 3-particle current, involves one resonant Andreev reflection
and one
reflection by the quasi-resonance at the energy gap edge, an equivalent
path 
exist also for injection from the left;
d: resonant 4-particle current involves two resonant Andreev reflections
(double resonance).}
\label{fig:mar}
\end{figure}

The single-particle current involves only a single crossing of the
tunnel 
barrier, schematically shown in \fig{mar}a. Thus, it cannot be
resonant, and the denominator $|z_{10}|^2$ can be considered in the 
lowest-order approximation.  The expression for the single particle
current, 
derived from Eqs.\ (\ref{Itotn}), (\ref{jX}), and (\ref{Iappn}) then
reads,
\begin{equation}
I_1={2Te\over \pi}\int_{\Delta-eV}^{-\Delta}dE\ N_L(E) N_R(E+eV)
\label{I1appr},
\end{equation}
where
\begin{eqnarray}
N_L(E)&=&{|E|\sqrt{E^2-\Delta^2}\over E^2-E_a^2} \label{Nl}\\
N_R(E)&=&{|E|\over\sqrt{E^2-\Delta^2} }
\end{eqnarray}
The form of \Eq{I1appr} is a standard one for a tunnel current given by
the 
tunnel Hamiltonian model were $N_R(E)$ is the density of states (DOS) in
the injection lead, while $N_L(E)$ has the meaning of the effective
phase dependent DOS in the interferometer. An important property of the
DOS of the interferometer is the disappearance of the singularities at
the gap edges for nonzero
phase difference. The reason is that the resonances (Andreev states)
move down
into the gap, according to \Eq{Al} (see \fig{dos}).
Consequently, the onset of the single-particle current at $eV=2\Delta$, 
which is of the order $T$ when $\phi=0$, is reduced at finite phase
difference (see \fig{iv2}).
%
%

In order to calculate the pair current we will first inspect the
corresponding transfer matrix,
$\hat{M}_{20}^{-1}=\hat{T}_{0}^{-1}\hat{U}_{1}^{-1}\hat{T}_{1}^{-1}$.
For $\nu=1$ (injection from the right) this transfer matrix has zeros in
the limit $T=0$ at the energies of the Andreev bound states in the 
interferometer, $E_1=\pm E_a$. The expansion of the matrix in powers of
$T$ has the form
\begin{eqnarray}
\hat{M}_{20}^{-1}&=&{1\over T|\xi_1|\cos{\phi\over
2}}\bigl[(E_1^2-E_a^2)(\sigma_z-i\sigma_y)(1-T) \nonumber \\
 &+& iT(E_1|\xi_1|-\sigma_x\sqrt{DR}\sin^2{\phi\over 2})\bigr]+ O(T^2),
\label{M20}
\end{eqnarray}
where we have used the identity, $\sin\theta=2\sqrt{DR}$, derived from
the unitarity condition \Eq{unit} and $\theta=\theta'+\rho$, in the 
limit $T=0$. Equation (\ref{M20}) clearly demonstrates the resonant
behavior. The first term is real and proportional to the deviation of
the energy $E_1$ for the Andreev reflection from the energy of the
Andreev state. This term determines the position of the resonance. The
second, imaginary, term is proportional to $T$ and determines the
width of the resonance. Both the position and width of the resonance
depend on the phase difference.
At $T\ll 1$ the spectral current is dominated by the Andreev resonance,
and neglecting the non-resonant current we arrive at the expression
\begin{equation}
I_{2}^R =  {2e\over \pi}\int_{\Delta-eV}^{eV-\Delta} dE_1{ \Gamma_+
\Gamma_-\over (E_1-E_a)^2+({\Gamma_+ + \Gamma_-  \over \displaystyle
2})^2}
\label{I2appr},
\end{equation}
where 
\begin{eqnarray}
\Gamma_{\pm} &=& TD_{\pm} N_R(E_a\mp eV), \\
D_{\pm} &=&
{1\over2}\left(\sqrt{\Delta^2-E_a^2}\pm{\sqrt{DR}\Delta^2\over
E_a}\sin^2{\phi\over 2}\right), 
\label{g2}
\end{eqnarray}
$0\leq D_{\pm}\leq \sqrt{D}$.
A factor of 2 appears in the current expression, in addition to the 
two-particle factor $n=2$ in \Eq{jX}, since there are two Andreev
resonances,
at $E_a$ and $-E_a$. The integrand in \Eq{I2appr} has the form of the
transmission coefficient for an asymmetric quantum mechanical double
barrier structure.
The difference in the tunnel rates $\Gamma_\pm$ is due to the different
DOS 
for the injection and exit energies, but also due to the different
effective transparencies of the Y-branch splitter, $TD_{\pm}$, for
electrons and holes. 
The resonant transparency of the junction for the pair current is
proportional, according to \Eq{I2appr}, to the product
\begin{equation}
D_+D_-={\Delta^2(\Delta^2-E_a^2)\over 4E_a^2}\cos^2{\phi\over 2},
\end{equation}
which turns to zero at $\phi=\pi$. Thus the resonant pair current is 
blocked at $\phi=\pi$, in agreement with our earlier general conclusion 
about the suppression of Andreev reflection at the interferometer in
this 
case. Evaluation of the integral over energy in \Eq{I2appr} 
under the condition, $E_a/\Delta<1-T$, yields the pair
current
\begin{equation}
I_{2}^R= {4e\Gamma_+\Gamma_-\over |\Gamma_++ \Gamma_-|}.
\label{I2a}
\end{equation}
The magnitude of this current is generally of order $T$; however, in the 
vicinity of the resonant onset, $eV=\Delta+E_a+e\delta V$ for deviations 
$e\delta V$, the rate 
$\Gamma_+$ increases due to the enhancement of the DOS in the right
superconductor, 
$\Gamma_+\sim1/\sqrt{\delta V}$, which yields the overshoot seen 
in Fig. \ref{fig:iv2}.  At lower voltage,
$\Delta<eV<\Delta+E_a$, the pair current is non-resonant and
proportional 
to $T^2$, which is also the case for the pair current $I_2^L$ injected
from the interferometer.
%
%

The resonant three-particle current can be analyzed in a similar
manner. We construct the corresponding transfer matrix using the pair
current matrix, \Eq{M20}, and for $\nu=1$ we get
$\hat{M}_{30}^{-1}=\hat{M}_{20}^{-1}\hat{U}_2^{-1}\hat{T}_2^{-1}$. 
This equation has resonances at the same energies as the pair current, 
$E_1=\pm E_a$, but nevertheless the resonant current is small, $I_3\sim
T^2$,
since the MAR path, \fig{mar}c, includes an additional crossing of the 
tunnel barrier. 
The magnitude of the current is considerably enhanced in the vicinity of 
voltage $eV=E_a+\Delta$ when the
second Andreev reflection occurs at the gap edge of the right electrode, 
$E_2=\Delta$  (see Fig.\ \ref{fig:mar}c). The singular DOS at the gap
edge in this
electrode appears in the MAR calculation as a quasibound state situated
at the gap edge. Thus, this MAR process can be
interpreted as an overlap of two resonances, one at the Andreev bound
state at
energy $-E_a$, and another at the quasibound state at the gap edge. A
similar
analysis applies to the current injected from the interferometer,
$\nu=-1$: in
this case, the double resonance is formed by the Andreev state at $E_a$
and
the quasibound state at lower gap edge. The height of
the double resonance peak is $(I_3)_{max}\sim T^{4/3}$.

A double resonance involving two Andreev bound states occurs in the
four-particle current injected from the right as shown in \fig{mar}d.
The corresponding MAR path is selected by the conditions $E_1=-E_a$ and
$E_3=E_a$.
The resulting peak is situated at $eV=E_a$ and has the hight
$(I_4)_{max}\sim T$.
In the vicinity of the peak, $e\delta V=eV-E_a\ll\Delta$, the 
expression for the resonant four-particle current becomes,
\begin{equation}
I_{4}^R = {e\over 2\pi}\int_{-\infty}^{\infty}{dE_2\
{\tilde \Gamma}_-^2\Gamma_0^2\over \left[E_2^2-(e\delta V)^2-{1\over
4}({\tilde \Gamma}_-^2+\Gamma_0^2)\right]^2+E_2^2{\tilde \Gamma}_-^2},
\label{I4}
\end{equation}
where
\begin{equation}
\Gamma_0 = TD_+ ,  \;\;\; {\tilde \Gamma}_- = TD_-N_R(2E_a).
\end{equation}
The current peak under consideration is a full-scale current structure
in a 
region of the IVC which in point contacts is dominated by non-resonant 
four-particle current, which is of the order $T^4$.
As any Andreev resonance, the four-particle current peak strongly
depends on the
phase difference; it disappears at $\phi=\pi$. The peak exists within
the
voltage interval $\Delta/2<eV<\Delta$ and within the phase difference
window 
$0<\sin^2\phi/2<3/(4D)$. 

We note that the positions of all discussed current structures are
proportional to
the order parameter $\Delta$ and therefore temperature dependent and
scale with $\Delta$.
%
%

We end this section with the discussion of CVC at large voltage, $eV\gg
2\Delta$. The current at large voltage results exclusively from the
single particle and pair currents. It is straightforward to calculate
the single
particle current at large voltage from \Eq{I1appr},
\begin{eqnarray}
I_1= {2T\over \pi}\int_{\Delta}^{eV-\Delta} {dE_1 E_1\xi_1E_0\over
(E_a^2-E_1^2)\xi_0} ={2Te^2V\over \pi}+I_{1,exc}.
\end{eqnarray}
The first term in this equation is the current through the normal
junction,
while the second term is the single-particle contribution to
the excess
current,
\begin{equation}
I_{1,exc}=-eT\sqrt{D}\Delta|\sin(\phi/2)| + O(T^2).
\end{equation}
This excess current is large, $\sim T$, and negative, which reflects the
suppression
of the current onset at $eV=2\Delta$. 
The pair current contribution to the excess current is dominated by the
resonance and is given by \Eq{I2a}. The total excess current to
leading order in $T$ has the form
\begin{equation}
I_{exc}=-eT\Delta\sqrt{D}R{|\sin(\phi/2)|^3\over
1-D\sin^2(\phi/2)}.
\end{equation}
The excess current in the interferometer is large and negative, in sharp
contrast 
to quantum point contacts, where the excess current is small, $\sim
T^2$, and
positive.\cite{Bratus1997} However, it turns to zero at $\phi=0$, in
agreement
with our observation that in this case the interferometer behaves as a
quantum
point contact. The excess current achieves its minimum value at
$\phi=\pi$
because
of blockade of the pair current.

\begin{figure}[b]
\psfig{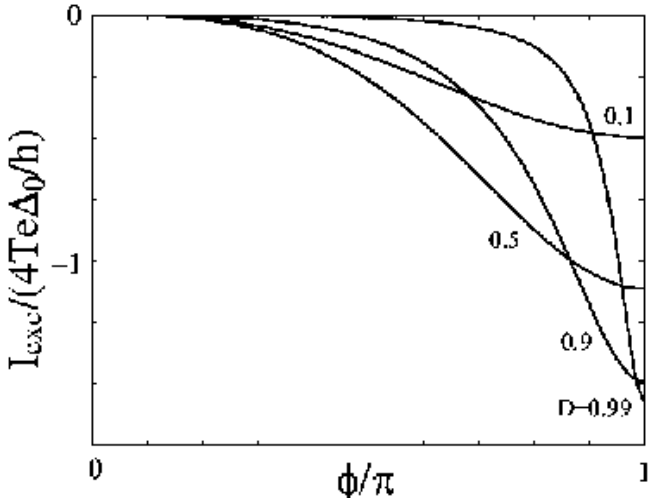} 
\caption{The first order ($\sim T$) excess current as a function of the 
phase difference for different transparencies $D$ of the SNS junction of
the
interferometer.} \label{fig:exc}
\end{figure}
%
\section{Conclusions}

We have investigated the phase dependence of the dc current-voltage
characteristics  (CVC) in SNS interferometers consisting of two
superconducting
reservoirs connected by a Y-shaped normally conducting quantum wire.
We have developed a generalization of the existing coherent MAR theory,
which
fully incorporates the effect of interference, as well as resonance
effects due to Andreev bound states in the interferometer. 

The phase dependent current-voltage characteristic $I(V,\phi)$ has been
numerically investigated for single-mode wires with length smaller than
the
superconducting
coherence length. The analytical study of the current was
performed using a resonance approximation in the limit of weak coupling
of the interferometer to the injection lead. 

We found significant enhancement of
the subgap current at finite phase difference compared to the case of
zero phase difference. The enhancement is accompanied
by a shift of the onset of the current at $2\Delta$ to smaller
voltages,
$\Delta+E_a(\phi)<2\Delta$, and appearance of current peaks at
$E_a$ and
$\Delta+E_a$.
These features are produced by resonant Andreev reflection of the
injected
particles by the S$_1$NS$_2$ junction of the interferometer:
they are most pronounced for weak coupling of the S$_1$NS$_2$ junction 
to the injection lead. The positions of all current structures are
temperature
dependent and scale with $\Delta$.

We further found strong suppression of the
current when the phase difference approaches $\phi\rightarrow \pi$: in
this region all subgap
current structures disappear and, moreover, the excess current at large
voltage becomes negative (deficiency current). The magnitude of the
deficiency
current is considerably large at $\phi=\pi$.
This effect results from the suppression of the Andreev reflections from
the  S$_1$N and  S$_2$N interfaces due to the 
interference in the arms of the beam splitter (cf. a similar effect in
NS interferometers\cite{Nakano1993,Hekking1993}).

We conclude our discussion with some brief comments on the connection of the
studied model to more realistic experimental devices based on 2D electron
gas structures. In such devices, the normal region has a typical size
which exceeds the
superconducting coherence length and it contains several conducting electronic
modes. Furthermore, the NS interfaces are not perfectly transparent.
However, this effect is less important since the interface reflection
coefficient in practice can be rather small ($<0.2$).
Increasing the length of the arms of the interferometer will result
in the appearance of new phase dependent current peaks due to increasing 
number of Andreev states in the SNS junction of the interferometer 
(cf. Ref.~\onlinecite{Samuelsson}).
Similarly, increasing the length of the injection lead will result in
the
appearance of current resonances which do not depend on the phase
difference, 
due to $\phi$-independent superconducting bound states in this
lead.\cite{Ingerman}
In interferometers with multimode wires, the resonant
current peaks will be smeared; however, one should still expect some
phase dependent resonant features similar to the case of diffusive NS
interferometers
\cite{Hekking1993}. The interference effect leading to the large
deficient
current at large applied voltage will survive in multimode junctions.

\section{Acknowledgements}
The authors gratefully acknowledge support from NFR and KVA (Sweden) 
and NEDO (Japan).


\end{document}